\documentclass[12pt]{article}
\begin{document}

\begin{center}
{\LARGE \textbf{$d^3=0$, $d^2=0$ Differential calculi on certain
non-commutative (super) spaces.}}\\[0pt]\vspace{1cm}

\vspace{1cm}

{\large M. EL BAZ\footnote{{\large moreagl@yahoo.co.uk}}, A. EL HASSOUNI
\footnote{{\large lhassoun@fsr.ac.ma}}, Y. HASSOUNI\footnote{{\large %
y-hassou@fsr.ac.ma}}\\\ and E.H. ZAKKARI\footnote{{\large %
hzakkari@hotmail.com}}.}\\[0pt]\vspace{0.5cm}\vspace{0.5cm}

Laboratory of Theoretical Physics\\[0pt]PO BOX 1014, University Mohammed V %
\\[0pt]Rabat, Morocco.
\end{center}

\bigskip

\vspace {3cm}

\textbf{Abstract:} In this paper, we construct a covariant
differential calculus on a quantum plane with two-parametric
quantum group as a symmetry group. The two cases $d^2=0$ and
$d^3=0$ are completely established.

We also construct differential calculi $n=2$ and $n=3$ nilpotent
on super quantum spaces with one and two-parametric symmetry
quantum supergroup.

\vspace{2cm}


\newpage\ \

\section{Introduction:}

A non-commutative quantum (super) space \cite{manin1, manin2} is an unital,
associative algebra with a quantum (super) group as a symmetry group. These
objects \cite{drinfeld, jimbo} have enriched the arena of mathematics and
mathematical physics: they appear in the context of theory of knots and
braids \cite{zachos}, as well as in the study of Yang-Baxter equations \cite
{wadati}. Quantum (super) groups are deformations of the enveloping (super)
algebra of classical Lie groups in the sense that one recovers the classical
(super) commutator when the deformation parameters go to some particular
values. Usually the generators of a quantum (super) group are assumed to
commute with the non-commuting coordinates of the corresponding (super)
plane. As a consequence, the quantum (super) plane admits a quantum group as
a symmetry group with only one parameter: ($GL_q(1/1)$) $GL_q(2)$ \cite
{corrigan}. More generally, one can obtain a multiparametric quantum (super)
group, if one relaxes this property (commutation between space coordinates
and group generators), namely, $GL_{p,q}(1/1)$ \cite{tahri} and $GL_{p,q}(2)$
\cite{cho} respectively in the two dimensional quantum superplane and
quantum plane cases.

Many authors \cite{tahri, kobayachi, wess1, berezinski,
schirrmacher, coquereaux1, coquereaux2, salih1} have also studied
differential calculus with nilpotency $n=2$ on (super) spaces with
one or two-parameter (super) group as symmetry groups. An adequate
way leading to generalization of this ordinary differential
calculus arises from the graded differential algebra
\cite{kerner1, kerner2, kerner3, notre, salih2, salih3}. The
latter involves a complex parameter that satisfies some conditions
allowing to obtain a consistent generalized differential calculus.
The most important property of
this calculus is that the operator $"d"$ satisfies $\{d^n=0$ $/$ $d^l\neq 0$%
, $1\leq l\leq n-1\}$ and it contains as a consequence, not only
first differentials $dx^i$ $,$ $i=1...m,$ but involves also higher
order differentials $d^jx^i,$ $j=1...n-1.$

In this paper, we construct covariant differential calculus
$d^3=0$ on certain quantum (super) spaces with one or
two-parametric quantum group as a symmetry groups. We will show
that our differential calculus is covariant under the algebra with
a quantum group structure. The complex $j$, which appears in the
Leibniz rule, is a third-root of unity and will be an interesting
and non trivial aspect of the differential calculus that we will
introduce.

This paper is organized as follows:

In section $2$ we start by recalling the two-parameter quantum group acting
on a two-dimensional quantum plane. We also establish $n=2$ and $n=3$
covariant differential calculi on this space following $R.$ $Coquereaux$
approach \cite{coquereaux1, coquereaux2}. It will be noticed that some
modifications have been brought up to this approach in order to adapt it to
the two-parameter quantum group symmetry and the $n=3$ differential
calculus. In section $3,$ the same method will be applied to construct the $%
n=2$, $n=3$ covariant differential calculus on $1+1$-dimensional superspace
with one parameter quantum supergroup as a supersymmetry group. In section $%
4 $ we generalize the results of section $3$ by taking the two-parameter
quantum group acting covariantly on the superspace.

\section{Differential calculus on a two-parametric quantum plane.}

\subsection{ Preliminaries}

The two dimensional quantum plane is an associative algebra generated by two
non-commuting coordinates $x$ and $y$ \cite{manin1, manin2, coquereaux1}
satisfying the relation:

\begin{equation}
xy=q \, yx \; , \;\;\; q\neq 0,1 \;\;\;(q\in C).
\end{equation}

In order to have a two-parameter quantum group $GL_{p,q^{^{\prime }}}(2)$ as
a symmetry group of such a space\cite{cho}, one must assume that the
coordinates do not commute in general with elements defining this group. \\%
Indeed, for a generic element $T=\left(
\begin{array}{cc}
a & b \\
c & d
\end{array}
\right)$ of $GL_{p,q^{^{\prime }}}(2)$, the relations between the
matrix entries and the coordinates are assumed to be \cite{cho}:

\begin{equation}
\begin{array}{ccccl}
x \, a & = & q_{11}\;a\,x ~~~~~~~~~~~~ y\,a & = & q_{21}\;a\,y \cr x \, b & =
& q_{12}\;b\,x ~~~~~~~~~~~~ y\,b & = & q_{22}\;b\,y \cr x \, c & = &
q_{13}\;c\,x ~~~~~~~~~~~~ y\,c & = & q_{23}\;c\,y \cr x \, d & = &
q_{14}\;d\,x ~~~~~~~~~~~~ y\,d & = & q_{24}\;d\,y. \cr
\end{array}
\end{equation}

The coordinates $x$ and $y$ transform under $T$ and $^tT$ (transposed
matrix) as:

\[
\left(
\begin{tabular}{l}
$x$ \\
$y$%
\end{tabular}
\right) \stackrel{T}{\longrightarrow }\left(
\begin{tabular}{l}
$x^{\prime }$ \\
$y^{\prime }$%
\end{tabular}
\right) =\left(
\begin{array}{cc}
a & b \\
c & d
\end{array}
\right) \left(
\begin{tabular}{l}
$x$ \\
$y$%
\end{tabular}
\right)
\]

\[
\left(
\begin{tabular}{l}
$x$ \\
$y$%
\end{tabular}
\right) \stackrel{^tT}{\longrightarrow }\left(
\begin{tabular}{l}
$x^{\prime \prime }$ \\
$y^{\prime \prime }$%
\end{tabular}
\right) =\left(
\begin{array}{cc}
a & c \\
b & d
\end{array}
\right) \left(
\begin{tabular}{l}
$x$ \\
$y$%
\end{tabular}
\right) .
\]

The requirement that the transformed coordinates obey a similar relation as $%
eq(1)$ (not necessarily with the same deformation parameter $q$) i.e.

\begin{eqnarray}
x^{\prime }y^{\prime } &=& \bar{q} \, y^{\prime }x^{\prime },\hspace{1.0in}\bar{%
q}\in C \\ x^{\prime \prime }y^{\prime \prime } &=& \;
\stackrel{=}{q}y^{\prime \prime }x^{\prime \prime
},\hspace{0.9in}\stackrel{=}{q}\in C,
\end{eqnarray}
and taking account of the defining relations of
$GL_{p,q^{\prime}}(2)$ \cite {schirrmacher},

\begin{eqnarray}
ab=p \, ba\hspace{1.0in}cd= &&p \, dc  \nonumber \\
ac=q^{^{\prime }}\, ca\hspace{1.0in}bd = &&q^{^{\prime }} \, db \\
p \, bc=q^{^{\prime }} \, cb\hspace{0.6in}ad-da= &&(p-\frac 1{q^{^{\prime
}}}) \, bc,  \nonumber
\end{eqnarray}
for some non zero $p$, $q^{^{\prime }}$ with $pq^{^{\prime }}\neq -1,$
implicates further constraints on the involved parameters:

\[
\bar{q} = \stackrel{=}{q}
\]
and
\begin{eqnarray}
q_{11}=1\hspace{1.2in}q_{21}&= &qq^{^{\prime }-1}k  \nonumber \\
q_{12}=\bar{q}p^{-1}\hspace{1.0in}q_{22}&=&q\bar{q}p^{-1}[\bar{q}%
-(p-q^{^{\prime }-1})k]  \nonumber \\
q_{13}=\bar{q}q^{^{\prime }-1}\hspace{1.0in}q_{23}&= &q\bar{q}%
q^{^{\prime }-1}[\bar{q}-(p-q^{^{\prime }-1})k] \\
q_{14}=\bar{q}q^{^{\prime }-1}k\hspace{1.0in}q_{24}&= &q\bar{q}%
^2q^{^{\prime }-1}p^{-1}[\bar{q}-(p-q^{^{\prime }-1})k].  \nonumber
\end{eqnarray}

One can check that the matrix $T=\left(
\begin{array}{cc}
a & b \\
c & d
\end{array}
\right) $ is indeed an element of the quantum group $GL_{p,q^{^{\prime
}}}(2) $ and is consistent with Hopf algebra structures \cite{cho}. For
supplementary properties and results concerning the quantum group $%
GL_{p,q^{^{\prime }}}(2)$ see for example \cite{schirrmacher}.

It is clear that many quantum planes could be associated to this
two-parameter quantum group, depending on choices of the $q_{ij}$'s. In the
following, we shall confine our selves to the case $\bar{q}=q$,
which corresponds to the standard definition of the quantum plane.

\subsection{Differential calculus with nilpotency $n=2$ $(d^2=0)$.}

Our aim in this section, is to construct a differential calculus on the
previously defined quantum plane. We proceed using the same approach as the
one adopted in \cite{coquereaux1, coquereaux2, notre}.

We start by defining the exterior differential $"d"$ which satisfies the
usual properties, namely:

$i/$ Linearity

$ii/$Nilpotency
\[
d^2=0.
\]
$iii/$ Leibniz rule
\[
d(uv) = d(u) v+(-1)^n ud(v),
\]
where $u\in \Omega ^n.$ $\Omega ^n$ is the space of forms with degree $n,$

\[
d:\Omega ^n\rightarrow \Omega ^{n+1}.
\]

$\Omega ^0$ is the algebra of functions defined on the quantum plane. We
have also:

\begin{equation}
d(x)= dx,\;\; d(y)= dy \;\; \hbox{and} \;\; d(1)=0.
\end{equation}

From (2), we deduce:

\begin{equation}
\begin{array}{ccccl}
(dx)a & = & q_{11} \, a(dx)~~~~~~~~~~~~(dy)a & = & q_{21} \, a(dy) \cr (dx)b &
= & q_{12}\, b(dx)~~~~~~~~~~~~(dy)b & = & q_{22} \, b(dy) \cr (dx)c & = &
q_{13} \,c(dx)~~~~~~~~~~~~(dy)c & = & q_{23} \, c(dy) \cr (dx)d & = & q_{14}
\, d(dx)~~~~~~~~~~~~(dy)d & = & q_{24} \, d(dy). \cr
\end{array}
\end{equation}

One can write \textit{a priori} $xdx,xdy,ydx$, and $ydy$ in terms of $%
(dx)x,(dy)x,(dx)y$ and $(dy)y$, by mean of $16$ unknown coefficients \cite
{coquereaux1}. Imposing the covariance of the obtained relations under $%
GL_{p,q^{^{\prime }}}(2)$, and differentiating $eq(1)$, permit to fix 15 of
the 16 unknown coefficients. The associativity of the expression $%
(xdx)dy=x(dxdy)$ enables us to fix the last unknown parameter.

We notice that in the usual case $Gl_q(2)$ this approach yields directly the
desired differential calculus. However, when $GL_{p,q^{^{\prime }}}(2)$ is a
symmetry group, we obtain additional conditions on the parameters $k$ and $%
q^{^{\prime }}$:

\begin{equation}
q^{^{\prime }}=q\;,\;k=\frac{q^{^{\prime }}}p.
\end{equation}

Then $eq(6)$ becomes:

\begin{equation}
q_{11}=q_{13}=1\;,\;q_{12}=q_{14}=q_{21}=q_{23}=qp^{-1}\;,%
\;q_{22}=q_{24}=q^2p^{-2}.
\end{equation}

So, the covariant differential calculus is given by:

\begin{eqnarray}
x \, dx &=& \frac 1{pq} \; dx \, x ~~~~~~~~~~~~~~~~~~ x \, dy
=\frac {1}{p} \; dy\, x  \nonumber \\ y\, dy &=& \frac {1}{pq}\;
dy\, y ~~~~~~~~~~~~~~~~~~ y\, dx =(\frac {1}{pq} - 1) \; dy\, x+
\frac {1}{q}\; dx\, y \\ dx\, dy &=& -\frac {1}{p} \; dy\, dx
~~~~~~~~~~~~~~ (dx)^2 = (dy)^2=0, \nonumber
\end{eqnarray}
and the differential algebra is $\Omega _{x,y}^{q,p}=\{x,$ $y,$ $dx,$ $dy\}.$

It is remarkable that the differential calculus on the quantum plane with
$GL_q(2)$, as a symmetry group \cite{wess1,berezinski,coquereaux1}, can be
obtained from the two-parameter one $eq(11)$ in the $p\rightarrow q$ limit.

As in the ordinary case, the differential operator $d$ can be realized by
\[
d := dx \, \partial x + dy \, \partial y \, .
\]
Based on this realization one can construct a gauge field theory on the
two-parameter quantum plane. This should be achieved formally as in \cite
{notre}.

The nilpotent differential calculus can be extended to higher orders, as
there is no reason to constrain this one to $n=2$ nilpotency \cite
{kerner1,kerner2,kerner3,notre,coquereaux3,dubois, dubois1, dubois2}.

In the following section, we generalize the differential calculus on the
quantum plane with one-parameter symmetry group \cite{notre} to the
two-parameter one, this is done by extending the $n=2$ differential calculus
obtained here to $n=3$ case.

\subsection{Differential calculus with nilpotency $n=3$ $(d^3=0).$}

Let us introduce the differential operator ''$d$'' that satisfies the
following conditions:

$i/$ Linearity

$ii/$ Nilpotency
\[
d^3=0\;,\;d^2\neq 0.
\]

$iii/$ Leibniz rule
\[
d(uv)=(du)v+(j)^nud(v).
\]
where $j$ is the cubic root of unity: $j=e^{\frac{2i\pi
}3},1+j+j^2=0$. $u$ is an element of $\Omega ^n$, the space of forms with degree $n$. It is a
subspace of the differential algebra $\tilde{\Omega} _{x,y}^{q,p}=%
\{x,y,dx,dy,d^2x,d^2y\}.$ The new objects $d^2x$ and $d^2y$ which appear are
defined by:

\[
d(dx)=d^2(x)=d^2x\;\;,\;\;d(dy)=d^2(y)=d^2y,
\]
these are ''forms'' with degree two.

In order to ensure the covariance of the differential calculus under the
two-parameter symmetry group $GL_{p,q^{^{\prime }}}(2$), we proceed as in
the previous section. However, instead of the last step where we have used
the associativity property, we shall use the independence between the two
different $2$-forms $z$ $d^2z^{^{\prime }}$ and $dz$ $dz^{^{\prime }}$,
where $z,$ $z^{^{\prime }}=x,$ $y.$ Below, we will discuss how to recover
this property.

The same constraints on $q^{^{\prime }}$ and $k$ $eq(9)$ are recovered, thus
the $q_{ij}$ 's are the same as in $eq(10).$ The covariant differential
calculus is then given by:

\begin{eqnarray}
x\, dx &=& j^2 \; dx\, x ~~~~~~~~~~~~~~~~~~~ x\, dy = -\frac{jq}{1+qp} \;
dy\, x+\frac{j^2qp-1}{1+qp} \; dx\, y  \nonumber \\
y\, dy &=&j^2 \; dy\, y ~~~~~~~~~~~~~~~~~~~ y\, dx = \frac{j^2-qp}{1+qp} \;
dy\, x - \frac{jp}{1+qp} \; dx\, y  \nonumber \\
x\, d^2x &=& j^2 \; d^2x\, x ~~~~~~~~~~~~~~~~~ x\, d^2y = -\frac{jq}{1+qp}
\; d^2y \, x + \frac{j^2qp-1}{1+qp} \; d^2x \, y  \nonumber \\
y\, d^2y &=& j^2 \; d^2y\, y ~~~~~~~~~~~~~~~~~ y\, d^2x = \frac{j^2-qp}{1+qp}
\; d^2y\, x - \frac{jp}{1+qp}\; d^2x\, y \\
dx\, d^2x &=& j \; d^2x\, dx ~~~~~~~~~~~~~~~ dx\, d^2y = -\frac q{1+qp}\;
d^2y\, dx + \frac{jqp-j^2}{1+qp}\; d^2x\, dy  \nonumber \\
dy\, d^2y &=& j \; d^2y\, dy ~~~~~~~~~~~~~~~ dy\, d^2x = \frac{j-j^2qp}{1+qp}
\; d^2y\, dx - \frac p{1+qp} \; d^2x\, dy  \nonumber \\
dx\, dy &=& q \; dy\, dx ~~~~~~~~~~~~~~~ d^2x\, d^2y = q \; d^2y\, d^2x.
\nonumber
\end{eqnarray}

Moreover, a realization of $"d"$ in terms of partial derivatives:
\begin{equation}
d=dx\, \partial _x+dy\, \partial _y
\end{equation}
permits us to have $(dx)^3=(dy)^3=0$ \cite{notre}.

\vspace{0.5cm}

We note that the differential algebra $\tilde{\Omega} _{x,y}^{q,p}$, defined
above, is not associative. One can check this statement by first assuming
that this property (associativity) is preserved, then deriving some
inconsistent relations. Especially, one expects, due to this assumption, the
two expressions $(x\, dx)dy$ and $x(dx\, dy)$ to be equal. However, using
(12) and successively moving the parenthesis, one obtains two expressions
which are manifestly not equal, unless $pq=j^2$.

Thus, the differential algebra $\tilde{\Omega} _{x,y}^{q,p}$ is associative
only when $pq=j^2$, otherwise it is not.

Another associative 3-nilpotent differential algebra, for $pq=j$, can be constructed basing on the method already
mentioned in section (2.2), with a proper substitution of the differential
operator $d^2=0$ with the one $d^3=0$, $(d^2 \neq 0)$. It follows from this
method that the commutation relations between the coordinates and their
first order differentials are given (by the first ones) in (11). The first,
second and third differentiations of these relations give rise to the
remaining commutation relations between $x,\, y,\, dx,\, dy,\, d^2x$ and $%
d^2y$.

\vspace{0.5cm}

The results of \cite{notre} (i.e., differential calculus on a reduced
quantum plane respectively with $q^3=1$ and $q^N=1$) can be recovered as
limiting cases of the one obtained here (12); this is done by taking the
adequate limit $p\rightarrow q$ (respectively with $q^3=1$ and $q^N=1$).

It is also remarkable that the case $n=3$ differential calculus was applied
to introduce interesting \textit{''Higher order gauge theories''} \cite
{kerner2, kerner3, notre}. Indeed, an interesting manner to do this (in the
present case) is to pursue the same steps of \cite{notre}.

\vspace{0.5cm}

Another important question arises at this step is how to adapt the techniques
applied in subsections $(2.2)$ and $(2.3)$ to the quantum superplane. This will
be developed in the next section.

\section{Differential calculus on a one-parameter quantum superplane.}

\subsection{$n=2$ Differential calculus.}

The $1+1$ dimensional quantum superspace, in $Manin^{\prime }s$ approach
\cite{manin2, corrigan}, is an algebra generated by a bosonic and a
fermionic coordinate satisfying the relations:

\begin{eqnarray}
x \theta &=& q \, \theta x\;\;,\;\;q\neq 0,1 \\
\theta ^2&=&0.
\end{eqnarray}

In analogy with the quantum plane, a symmetry supergroup of this space is $%
GL_q(1/1)$, and a generic element of this supergroup is a supermatrix: $%
T=\left(
\begin{array}{cc}
a & \beta \\
\gamma & d
\end{array}
\right) ,$ where $a$, $d$ are bosonic elements commuting with $x$ and $%
\theta $ while $\beta ,\gamma $ are fermionic elements commuting with $x$,
anticommuting with $\theta $ and obeying the following relations:

\begin{equation}
\begin{array}{rclcc}
a\beta & = & q \, \beta a ~~~~~~~~~~~~~ d\beta & = & q \, \beta d \cr a\gamma
& = & q \, \gamma a ~~~~~~~~~~~~~ d\gamma & = & q \, \gamma d \cr \beta
\gamma +\gamma \beta & = & 0 ~~~~~~~~~~~~~~~~~ \beta ^2 & = & \gamma ^2=0 %
\cr ad-da & = & (q^{-1}-q)\, \beta \gamma . &  & \cr
\end{array}
\end{equation}

These relations can also be obtained by imposing the invariance of $%
eqs(14,15)$ under $T$ and $^{st}T$ $=\left(
\begin{array}{cc}
a & -\gamma \\
\beta & d
\end{array}
\right) $(supertranspose).

Many authors studied the differential calculus on this superspace \cite
{tahri, kobayachi,soni}. Here we construct the differential calculus based
on the same technique adopted by \textit{R. Couquereaux} \cite{coquereaux1} which is
used in the previous section, with however, some modifications to adapt it
to this superspace. We introduce an exterior differential operator $"d"$
satisfying the properties:

$i/$ Linearity
\begin{equation}
d(\lambda u)=(-1)^{\stackrel{\wedge }{\lambda }} \; \lambda \, d(u),
\end{equation}
where the parity $\stackrel{\wedge }{\lambda }=0,1$ respectively, if $%
\lambda $ is a bosonic or a fermionic element.

$ii/$ Nilpotency
\[
d^2=0.
\]

$iii/$Leibniz rule
\begin{equation}
d(uv)=(du)v+(-1)^{\stackrel{\wedge }{u}}(-1)^{\deg u}u(dv),
\end{equation}
where $\stackrel{\wedge }{u}$ is the parity of $u$ and $degu$ is the degree
of the differential form $u.$

Note that consistency requires that $d\theta $ commutes with $a$, $d$, $\beta
$, $\gamma $ and $dx$ commutes with $a$, $d$ and anticommutes with $\beta ,$ $%
\gamma .$

The same method applied in section $2.2$ yields:

\begin{eqnarray}
x\, dx &=& q^{-2} \; dx\, x ~~~~~~~~~~~~ x\, d\theta \; = \; q^{-1} \;
d\theta \, x  \nonumber \\
\theta \, d\theta &=& d\theta \, \theta ~~~~~~~~~~~~~~~~~~ \theta \, dx \; =
\; (1-q^{-2}) \; d\theta \, x- q^{-1} \; dx\, \theta \\
dx\, d\theta &=& q^{-1} \; d\theta \, dx ~~~~~~~~~ (dx)^2 \; = \; 0,
\nonumber
\end{eqnarray}
and the associative differential algebra is denoted $\Omega
_{x,\theta }^q=\{x,$ $\theta ,$ $dx,$ $d\theta \}.$

As in section $2.3$, one can apply the same method to generalize the
differential calculus on the superspace to higher orders ($d^3=0$). This is
the aim of the next section.

\subsection{ Differential calculus on superspace with nilpotency \protect\\ $%
n=3$ $(d^3=0).$}

We proceed as in section $2.3$, in order to construct the $n=3$ covariant
differential calculus on superspace. We introduce a differential operator $%
"d"$ satisfying the usual requirements, namely: linearity is the same as in $%
eq(17)$, the nilpotency will be changed to $n=3$ $(d^3=0)$ and the Leibniz
rule, $eq(18)$ becomes:
\begin{equation}
d(uv)=(du)v+(-1)^{\stackrel{\wedge }{u}}(j)^{\deg u}u(dv).
\end{equation}

The resulting differential algebra $\tilde{\Omega }_{x,\theta }^q$ is
generated by, $x,$ $\theta ,$ $dx,$ $d\theta ,$ $d^2x$ and $d^2\theta $
satisfying:
\begin{eqnarray}
x\, dx &=& j^2 \; dx\, x ~~~~~~~~~~~~~~~~~~~ x\, d\theta = -\frac{jq}{1+q^2}
\; d\theta \, x+\frac{j^2q^2-1}{1+q^2} \; dx \, \theta  \nonumber \\
\theta \, d\theta &=& d\theta \, \theta ~~~~~~~~~~~~~~~~~~~~~~~ \theta \, dx
= \frac{q^2-j^2}{1+q^2} \;d\theta \, x+\frac{jq}{1+q^2} \; dx \, \theta
\nonumber \\
dx\, d\theta &=& -q \; d\theta \, dx ~~~~~~~~~~~~~~~(d\theta )^2 = 0
\nonumber \\
x\, d^2x &=& j^2 \; d^2x\, x ~~~~~~~~~~~~~~~~\, x\, d^2\theta = -\frac{jq}{%
1+q^2}\; d^2\theta \, x+\frac{j^2q^2-1}{ 1+q^2} \; d^2x\, \theta  \nonumber
\\
\theta \, d^2\theta &=& -d^2\theta \, \theta ~~~~~~~~~~~~~~~~~~ \theta \,
d^2x = \frac{j^2-q^2}{1+q^2} \; d^2\theta \, x-\frac{jq}{1+q^2}\; d^2x\,
\theta \\
dx\, d^2x &=& j \; d^2x\, dx ~~~~~~~~~~~~~~~ dx\, d^2\theta = \frac q{1+q^2}
\; d^2\theta \, dx+\frac{jq^2-j^2}{1+q^2} \; d^2x\, d\theta  \nonumber \\
d\theta \, d^2\theta &=& j^2 \; d^2\theta \, d\theta ~~~~~~~~~~~~~~ d\theta
\, d^2x = \frac{j^2q^2-j}{1+q^2} \; d^2\theta \, dx - \frac q{1+q^2}\;
d^2x\, d\theta  \nonumber \\
d^2x\, d^2\theta &=& q \; d^2\theta \, d^2x ~~~~~~~~~~~~~~ (d^2\theta )^2 =
0.  \nonumber
\end{eqnarray}

Let us point out that the differential algebra $\tilde{\Omega
}_{x,\theta }^q $ is not associative, unless $q=j$. In the case
$q\neq j$, one can recover
this property by following the same steps mentioned at the end of section $%
2.3$ with the adequate modifications.

\section{Differential calculus on a two-parameter quantum superplane.}

\subsection{Differential calculus with nilpotency $n=2$, $(d^2=0)$.}

In this section, we generalize the results of section $3,$ in the sense that
we choose a two-parametric quantum supergroup $GL_{p,q^{\prime }}(1/1)$ as a
symmetry group for the superplane $eqs(14,15)$. This group will be
introduced using the same method as in section $2$ \cite{tahri,cho}.

The entries of a matrix element $T=\left(
\begin{array}{cc}
a & \beta \\
\gamma & d
\end{array}
\right) $ of $\, GL_{p,q^{\prime }}(1/1)$ satisfy the following non trivial
relations:

\begin{equation}
\begin{array}{rclcr}
a\beta & = & p\; \beta a ~~~~~~~~~~~~~ d\beta & = & p \; \beta d \cr a\gamma
& = & q^{^{\prime }} \; \gamma a ~~~~~~~~~~~~~ d\gamma & = & q^{^{\prime }}
\; \gamma d \cr p \; \beta \gamma + q^{^{\prime }} \; \gamma \beta & = & 0
~~~~~~~~~~~~~~~~~~~ \beta ^2 & = & \gamma ^2=0 \cr ad-da & = & (q^{^{\prime
}-1}-p) \; \beta \gamma . &  & \cr
\end{array}
\end{equation}

As it is done in section $2$ this superspace is covariant under $T$ and $%
^{st}T$ (supertranspose), and the analogous of $eq(2)$ are:
\begin{equation}
\begin{array}{cclcrcc}
x \, a & = & k \;a\,x & ~~~~~~~~~~~~ & \theta\,a & = & q\bar{q}q^{^{\prime
}-1}p^{-1}k \;a\,\theta \cr x \, b & = & \bar{q}p^{-1}k \;b\,x & ~~~~~~~~~~~~
& \theta\,b & = & - q\bar{q}^2q^{\prime -1}p^{-2}k \;b\,\theta \cr x \, c & =
& \bar{q}q^{^{\prime }-1}k \;c\,x & ~~~~~~~~~~~~ & \theta\,c & = & - q\bar{q}%
^2q^{^{\prime }-2}p^{-1}k \;c\,\theta \cr x \, d & = & \bar{q}^2q^{^{\prime
}-1}p^{-1}k \;d\,x & ~~~~~~~~~~~~ & \theta\,d & = & q \bar{q}^3q^{^{\prime
}-2}p^{-2}k \;d\,\theta. \cr
\end{array}
\end{equation}

We are interested in establishing a covariant differential calculus on this
superspace in the case $\bar{q}=$ $\stackrel{=}{q}$ $=$ $q.$
To achieve this construction, for $n=2$, we introduce a differential
operator ''$d"$ satisfying the same properties as in section $3.2$ (Linearity
$eq(17)$, nilpotency and Leibniz rule $eq(18)$). The associative differential algebra $%
\Omega _{x,\theta }^{p,q}=\{x,$ $\theta ,$ $dx,$ $d\theta \}$ is generated
by the following relations:
\begin{eqnarray}
x\, dx &=& (qp)^{-1}\; dx\, x ~~~~~~~~~~~~ x\, d\theta = p^{-1} \;
d\theta \, x  \nonumber \\ \theta \, d\theta &=& d\theta \, \theta
~~~~~~~~~~~~~~~~~~~~~\, \theta \, dx = (1-(qp)^{-1}) \; d\theta \,
x- q^{-1} \; dx\, \theta \\ dx\, d\theta &=& p^{-1} \; d\theta \,
dx ~~~~~~~~~~~~~ (dx)^2 = 0.  \nonumber
\end{eqnarray}

We have used $q^{\prime }=q$ and $k=\frac qp$, which, as in $eq(9)$, are
consequences of the requirement of the covariance of $\Omega _{x,\theta
}^{p,q}$ under $GL_{p,q^{\prime }}(1/1)$.

As expected, in the limit $p\rightarrow q$, we recover $\Omega _{x,\theta
}^q $ and relations (19).

\subsection{Differential calculus with nilpotency $n=3$ $(d^3=0).$}

The technique used in sections $(2.3)$ and $(3.3)$, allows us to construct
the $n=3$ differential algebra $\tilde{\Omega }_{x,\theta }^{p,q}=\{x,$ $%
\theta ,$ $dx,$ $d\theta ,$ $d^2x,$ $d^2\theta \}:$
\begin{eqnarray}
x\, dx &=& j^2 \; dx\, x ~~~~~~~~~~~~~~~~~~~ x\, d\theta =-\frac{jq}{1+qp}
\; d\theta \, x+\frac{j^2qp-1}{1+qp}\; dx \, \theta  \nonumber \\
\theta \, d\theta &=& d\theta \, \theta ~~~~~~~~~~~~~~~~~~~~~~~ \theta \, dx
=\frac{qp-j^2}{1+qp} \; d\theta \, x+\frac{jp}{1+qp} \; dx \, \theta
\nonumber \\
dx\, d\theta &=& -q \; d\theta \, dx ~~~~~~~~~~~~~~~ (d\theta )^2 = 0
\nonumber \\
x\, d^2x &=& j^2 \; d^2x\, x ~~~~~~~~~~~~~~~~ x\, d^2\theta =-\frac{jq}{1+qp}%
\; d^2\theta \, x+\frac{j^2qp-1}{1+qp}\; d^2x\, \theta
\\ \theta \, d^2\theta &=&-d^2\theta \, \theta ~~~~~~~~~~~~~~~~
\theta \, d^2x = \frac{j^2-qp}{1+qp} \; d^2\theta \,
x-\frac{jp}{1+qp} \; d^2x\, \theta \nonumber \\ dx\, d^2x &=& j \;
d^2x\, dx ~~~~~~~~~~~~~ dx\, d^2\theta = \frac q{1+qp}\; d^2\theta
\, dx+\frac{jqp-j^2}{1+qp}\; d^2x\, d\theta  \nonumber \\ d\theta
\, d^2\theta &=& j^2 \; d^2\theta \, d\theta ~~~~~~~~~~~~~ d\theta
\, d^2x = \frac{j^2qp-j}{1+qp}\; d^2\theta \, dx-\frac p{1+qp}\;
d^2x\, d\theta  \nonumber \\ d^2x\, d^2\theta &=& q \; d^2\theta
\, d^2x ~~~~~~~~~~~~~ (d^2\theta )^2=0. \nonumber
\end{eqnarray}

The same limit as in section $2.3,$ namely $p\rightarrow q,$ yields $\tilde{%
\Omega }_{x,\theta }^q$. The differential algebra $\tilde{\Omega
}_{x,\theta }^{p,q}$ is not associative. In order to restore this
property we proceed as mentioned at the end of sections $2.3$ and
$3.2$.

One physical application of the differential calculi (sections 3 and 4) is
to construct a supersymmetric gauge field theory on the quantum superplane
(with one or two parameter quantum supergroup as symmetry groups; the latter
will be a generalization of the former). However, this is not
straightforward, since one should firstly start by defining a supersymmetric
covariant derivative.

\section{Conclusion:}

In this paper, we have constructed differential calculi on certain quantum
(super) spaces. Namely, the $n=2$ and $n=3$ nilpotent differential calculi on
the quantum plane with two parametric quantum group ($GL_{p,q}(2)$) as a
symmetry group was obtained. We have also considered two cases of quantum
superplanes related to the one and two-parametric quantum supergroups $%
GL_q(1|1)$ and $GL_{p,q}(1|1)$, as symmetry groups, respectively. The
related $n=2$ and $n=3$ differential calculi were also established.

In general, the differential calculus can be applied to formulate gauge
field theories \cite{wess2, wess3, wess4}. As a consequence, the results
obtained here permit us to construct gauge theories on the corresponding
non-commutative spaces \cite{coming}. Indeed, for the quantum space (section
2), this can be done using the same techniques of \cite{notre}, where the
symmetry group is a one-parameter.

The non-commutative supersymmetric case (sections 3 and 4) will be treated
in the same fashion, with however, more care since it is essential first, to
define a covariant supersymmetric derivative \cite{nordine, west}.

We note that the differential calculus was also applied to derive a
corresponding quantum oscillator, where the latter is seen as a
representation of the former \cite{mishra}. It will be interesting to
achieve this with the differential calculus in section 2, as the resulting
quantum oscillator will be two-parameter dependent.

\end{document}